# EXPLICIT AND IMPLICIT FINITE DIFFERENCE SOLVERS IMPLEMENTED IN JAX FOR SHOCK WAVE PHYSICS

| **Avinash Potluri** | **Arturo Rodriguez** | **Taylor N. Garcia** | **Chelsea M. Caballero** |
|---|---|---|---|
| Texas A&M University | Texas A&M University | Texas A&M University | Texas A&M University |
| Kingsville, TX | Kingsville, TX | Kingsville, TX | Kingsville, TX |

| **Katrina I. Sanchez** | **Payal Helambe** | **Vineeth V. Kumar** | **Francisco O. Aguirre Ortega** |
|---|---|---|---|
| Texas A&M University | Cranfield University | Texas A&M University | Texas A&M University |
| Kingsville, TX | Bedford, UK | Kingsville, TX | Kingsville, TX |

## ABSTRACT

*Shock dynamics and nonlinear wave propagation are fundamental to computational fluid dynamics (CFD) and high-speed flow modeling. In this study, we developed explicit and implicit finite-difference solvers for the one-dimensional Burgers viscous equation to model shock formation, propagation, and dissipation. The governing equation, which incorporates convective and diffusive effects, serves as a simplified analogue of the Navier-Stokes equations. Using the Finite-JAX framework, each solver is implemented with upwind and central finite-difference schemes for the convective and diffusive terms, respectively. Time integration is performed using explicit forward Euler and implicit backward-time central space (BTCS) schemes under periodic and Dirichlet boundary conditions. Stability is ensured by the Courant-Friedrichs-Lewy (CFL) criteria for the convective and diffusive components. Numerical experiments quantify the accuracy, convergence, and real-time performance of JAX across CPUs, GPUs, and TPUs, demonstrating that JAX maintains fidelity while achieving portability. The results show that the explicit scheme captures impact accurately under strict time-step constraints, while the implicit formulation provides greater stability and accuracy at a higher computational cost. Taken together, these results establish a reproducible dataset for benchmarking CFD solvers and training machine learning models for nonlinear transport and impact-driven phenomena. Our new implementation of FiniteJAX enhances the portability, scalability, and performance of solvers based on the JAX framework developed by Google DeepMind.*

## NOMENCLATURE

| | |
|---|---|
| $u$ | Velocity field (m/s) |
| $x$ | Spatial coordinate (m) |
| $t$ | Time variable (s) |
| $\nu$ | kinematic viscosity (m²/s) |
| $\Delta x$ | Spatial grid spacing (m) |
| $\Delta t$ | Temporal step size (s) |
| $N$ | Number of spatial grid points |
| $i$ | Spatial index (grid node) |
| $n$ | Temporal index (time step) |
| $u_i^n$ | Discrete velocity at position (i), time step (n) |
| $L$ | Length of computational domain (m) |
| $CFL$ | Courant-Friedrichs-Lewy stability number |
| $BTCS$ | Backward-Time central space (implicit) |
| $RK4$ | Fourth-order Runge-Kutta time integration |
| $JAX$ | Google DeepMind High-Performance Library |
| $L2$ | Root mean square (RMS) error norm |
| $GPU$ | Graphical processing unit |
| $TPU$ | Tensor processing unit |
| $k$ | Runge-Kutta Steps |

## 1. INTRODUCTION

Shock waves play a fundamental role in the study of high-speed flows, aerodynamics, and compressible fluid mechanics. Their inherently nonlinear and dissipative nature poses challenges for both analytical modeling and numerical simulation, especially when steep gradients lead to discontinuities [1, 2]. Accurately capturing these discontinuities while maintaining numerical stability and physical fidelity remains a central goal of computational fluid dynamics (CFD). To investigate these nonlinear features in a tractable setting, the one-dimensional viscous Burgers' equation serves as a canonical model that encapsulates the essential physics of convection, advection, diffusion, and shock formation [3, 4]. Despite its relative simplicity, Burgers' equation captures the fundamental interplay between nonlinear advection and viscous dissipation that governs shock evolution in more complex flow systems [5].

Finite-difference methods have long been employed to discretize and solve partial differential equations in CFD. Among these, explicit schemes offer simplicity and computational efficiency. Still, they are constrained by stringent stability limits, whereas implicit schemes provide improved stability for stiff problems at the expense of increased computational cost [6-11]. The balance between these characteristics offers valuable insight into numerical trade-offs that influence solver design for both classical and modern applications. In addition to finite-difference approaches, spectral methods have been established for high-order accuracy in smooth flow regimes. In contrast, finite element (FEM) and finite volume (FVM) methods have become foundational tools for complex geometries and conservation-based formulations [12-16]. More recently, deep learning–based solvers, such as Physics-Informed Neural Networks (PINNs) and Partition of Unity PINNs (POU-PINNs), have emerged as powerful alternatives that embed physical laws directly into neural architectures [17-23]. By systematically implementing and comparing explicit and implicit formulations, one can evaluate the stability, accuracy, and efficiency that define high-performance fluid flow solvers.

In recent years, advances in differentiable programming and high-performance computing have reshaped scientific computing. Frameworks such as JAX, developed by Google DeepMind, provide automatic differentiation and just-in-time compilation, enabling numerical solvers to run efficiently on CPUs, GPUs, and TPUs [24-30]. The Finite-JAX framework extends traditional finite-difference schemes within this paradigm, combining numerical rigor with computational scalability [31, 33]. This integration offers a new opportunity to unify physics-based solvers with machine learning workflows, enabling both forward simulations and gradient-based optimization [34].

The objective of this study is to develop explicit and implicit finite-difference solvers for the one-dimensional viscous Burgers' equation using the Finite-JAX platform, to generate a reproducible dataset for shock dynamics research. The numerical schemes employ upwind differencing for the advective term and central differencing for the diffusive term. The upwind scheme, though lower in order, is chosen for the convective term due to its inherent numerical stability and monotonicity, which are essential for capturing steep gradients and shock fronts without introducing spurious oscillations. In contrast, the central difference scheme, being second-order accurate, is applied to the diffusive term to achieve higher precision in modeling viscous smoothing effects, resulting in smoother solution gradients and less restrictive stability requirements. The stability of the solvers is assessed using the Courant–Friedrichs–Lewy (CFL) conditions for advection, convection, and diffusion. Numerical results are compared for accuracy, convergence, and computational wall time across different hardware architectures. This approach not only benchmarks the performance of differentiable solvers but also contributes to a growing foundation for hybrid CFD–AI research, where numerical accuracy and data-driven modeling intersect.

## 2. MATERIALS AND METHODS

### 2.1 Governing Equation

The governing equation for the present study is the one-dimensional viscous Burgers' equation, expressed as:

$$\frac{\partial u}{\partial t} + u \frac{\partial u}{\partial x} = \nu \frac{\partial^2 u}{\partial x^2} \qquad (1)$$

where $u(x,t)$ represents the velocity field, and $\nu$ it is the kinematic viscosity (or the diffusion coefficient). This equation serves as a nonlinear partial differential equation that models both convective transport and diffusive dissipation. The balance between these two terms gives rise to steep gradients and shock-like structures that mimic compressible flow behavior in a simplified form.

### 2.2 Spatial and Temporal Discretization

The computational domain is discretized into $N$ equally spaced grid points along the spatial direction, $x_i = i\Delta x$, where $i = 0,1,\dots,N$. Time is advanced in discrete intervals. $t^n = n\Delta t$, where $n = 0,1,2,\dots$. The speed at each grid point and time step is denoted as $u_i^n \approx u(x_i, t^n)$.

For temporal evolution, a forward difference approximation is applied to the time derivative:

$$\frac{\partial u}{\partial t} \approx \frac{u_i^{n+1} - u_i^n}{\Delta t} \qquad (2)$$

Spatial derivatives are approximated using finite differences as follows:

- For the convective term, the backward difference is used for stability

$$\frac{\partial u}{\partial x} \approx \frac{u_i^n - u_{i-1}^n}{\Delta x} \qquad (3)$$

In the Burgers' equation, the convective flux term $\frac{\partial u}{\partial x}$. It becomes nonlinear when multiplied by u, since the velocity field determines the direction of propagation. When this term is discretized as $u_i^n \frac{u_i^n - u_{i-1}^n}{\Delta x}$, advective term. The multiplication by $u_i^n$ effectively selects the upwind direction according to the local flow velocity. If $u_i^n > 0$, information propagates rightward, and the backward difference $(u_i^n - u^n_{i-1})$ is used; if $u_i^n < 0$, a forward difference $(u^n_{i+1} - u_i^n)$ would be applied instead. This self-adaptive property makes the upwind scheme naturally suited for handling nonlinear advection and preventing non-physical oscillations near sharp gradients.

- For the diffusive term, a second-order central difference scheme is used:

$$\frac{\partial^2 u}{\partial x^2} = \frac{u^n_{i+1} - 2u_i^n + u^n_{i-1}}{\Delta x^2} \quad (4)$$

Combining these approximations yields the explicit finite-difference scheme:

$$u_i^{n+1} = u_i^n - \frac{\Delta t}{\Delta x} u_i^n (u_i^n - u^n_{i-1}) + \frac{\nu \Delta t}{\Delta x^2}(u^n_{i+1} - 2u_i^n - u^n_{i-1}) \quad (5)$$

For the implicit solver, the convective and diffusive terms are evaluated at the next step. $n + 1$, forming a tridiagonal system that is solved iteratively using backward-time central-space (BTCS) or Crank–Nicolson (Implicit Scheme) schemes. These implicit formulations allow for larger time steps while maintaining numerical stability.

**2.3 Explicit Finite-Difference Scheme (Forward-Euler Upwind–Central Method)**

For the explicit solver, the time derivative is discretized using a forward difference, the convection term with a first-order upwind scheme for stability, and the diffusion term with a second-order central difference.

The resulting discrete scheme is:

$$u_i^{n+1} = u_i^n - \frac{\Delta t}{\Delta x} u_i^n (u_i^n - u^n_{i-1}) + \frac{\nu \Delta t}{\Delta x^2}(u^n_{i+1} - 2u_i^n - u^n_{i-1}) \quad (6)$$

This scheme is explicit in time and advances. $u$ directly from $t$ to $t^{n+1}$. It is first-order accurate in time and second-order accurate in space.

Stability constraint (CFL conditions):

$$\frac{u_{max} \Delta t}{\Delta x} \leq 1, \quad (7)$$

$$\frac{\nu \Delta t}{\Delta x^2} \leq \frac{1}{2}, \quad (8)$$

These ensure that information propagates within one spatial grid per timestep (convective stability) and that the diffusion term does not overshoot (diffusive stability).

**2.4 Implicit Finite-Difference Scheme (Backward-Time Central-Space, BTCS)**

The implicit form of the Burgers' equation is obtained by evaluating all spatial derivatives at the future timestep. $t^{n+1}$.

$$u_i^{n+1} = u_i^n - \frac{\Delta t}{\Delta x} u_i^n (u_i^n - u^n_{i-1}) + \frac{\nu \Delta t}{\Delta x^2}(u^n_{i+1} - 2u_i^n - u^n_{i-1}) \quad (9)$$

This system is solved iteratively using Gauss–Seidel or Newton–Raphson iteration until convergence at each timestep.

The implicit scheme is unconditionally stable for linear problems and is robust for stiff systems, though it is computationally more expensive.

Jacobian Formulation:

In the matrix form, the system can be represented as:

$$F(u^{n+1}) = 0 \quad (10)$$

Where F is the residual vector:

$$F_i = u_i^{n+1} - u_i^n + \Delta t \left[ u_i^{n+1} \frac{u^{n+1}_{i+1} - u^{n+1}_{i-1}}{2\Delta x} - \nu \frac{u^{n+1}_{i+1} - 2u_i^{n+1} + u^{n+1}_{i-1}}{\Delta x^2} \right] \quad (11)$$

To solve this nonlinear system, a Newton-Raphson iteration is applied:

$$J(u^{n+1,k})\delta u = -F(u^{n+1,k}) \quad (12)$$

$$u^{n+1,k+1} = u^{n+1,k} + \delta u \quad (13)$$

Here, $J$ is the Jacobian matrix, defined as the derivative of the residual vector with respect to the unknown:

$$J_{ij} = \frac{\partial F_i}{\partial u_j^{n+1}} \quad (14)$$

For the one-dimensional Burgers' equation, $J$ is tridiagonal, since each residual $F_i$ depends only on $u^{n+1}_{i-1}$, $u_i^{n+1}$ and $u^{n+1}_{i+1}$.

The entries of Jacobian are given by:

$$J_{i,j-1} = -\frac{\Delta t}{2\Delta x} u^{n+1}_i + \frac{\nu \Delta t}{\Delta x^2} \quad (15)$$

## 2.5 Boundary and Initial Conditions

**Boundary Conditions**

Two types of boundary conditions were used depending on the case:

- Dirichlet (Fixed Value):

$$u(0,t) = a, \quad (16)$$
$$u(L,t) = b \quad (17)$$

Used to enforce specific inflow and outflow conditions.

- Periodic:

$$u_0^n = u_N^n \quad (18)$$

Used to simulate cyclic or continuous shock propagation across the domain.

**Initial Condition**

- Sinusoidal:

$$u(x,0) = \sin(\pi x) \quad (19)$$

- Step Function (Shock Formation):

$$u(x,0) = \begin{cases} 1, & x < 0.5L \\ 0, & x \geq 0.5L \end{cases} \quad (20)$$

The step condition generates a shock front that evolves due to the competing effects of nonlinear advection and viscous diffusion.

## 2.6 Computational Setup

Both explicit and implicit solvers were implemented using standard Python and NumPy without JAX acceleration.

The simulations were executed on a single-core CPU (Intel i7, 3.2 GHz), with uniform grids of $N = 201$ and timesteps chosen to satisfy the CFL criteria.

The simulation outputs include temporal velocity profiles, contour plots, and $L_2$-norm error comparisons against analytical or steady-state reference solutions obtained via the Cole–Hopf transformation. The Cole-Hopf transformation linearizes the advection term when applied between spaces [5].

## 2.7 Finite-JAX Solver Implementation

To extend the finite-difference formulation into a differentiable computing environment, both explicit and implicit solvers were reimplemented using the Finite-JAX framework. Finite-JAX is built upon Google's JAX library, which provides automatic differentiation, just-in-time (JIT) compilation, and device-agnostic execution across Central Processing Units (CPUs), Graphics Processing Units (GPUs), and Tensor Processing Units (TPUs).

The JAX environment allows efficient vectorization through the XLA (Accelerated Linear Algebra) compiler, transforming standard Python functions into optimized machine code. This enables CFD solvers to achieve significant speedups while maintaining numerical transparency and precision.

### 2.7.1 Explicit Solver (JAX–Euler and RK4)

For the explicit formulation, the solver employs the exact spatial discretization as the non-JAX implementation:

$$u_i^{n+1} = u_i^n - \frac{\Delta t}{\Delta x} u_i^n (u_i^n - u_{i-1}^n) + \frac{\nu \Delta t}{\Delta x^2} (u_{i+1}^n - 2u_i^n + u_{i-1}^n) \quad (21)$$

However, in JAX, computations are defined as pure functions and compiled with `jax.jit`, which fuses operations into a single kernel, minimizing memory overhead.

Two time-integration strategies were implemented:

- Forward Euler (First-Order): identical to the explicit update above, providing high efficiency for short-time transient simulations.
- Runge–Kutta Four (Fourth Order): improves temporal accuracy, defined as:

$$u_i^{n+1} = u^n + \frac{\Delta t}{6}(k_1 + 2k_2 + 2k_3 + k_4) \quad (22)$$

where each $k_j$ represents intermediate stages computed using the same Burgers' equation operator.

The explicit JAX solvers are particularly suited for performance benchmarking and differentiable data generation, as gradients of the numerical solution with respect to initial or boundary conditions can be computed automatically via reverse-mode auto differentiation.

### 2.7.2 Implicit Solver (JAX–BTCS)

The implicit solver extends the Backward-Time Central-Space (BTCS) method by utilizing JAX linear algebra primitives, such as `jax.numpy.linalg.solve`.

At each time step, a tridiagonal system is assembled as:

$$A_{ij} u_i^{n+1} = u^n \quad (23)$$

where the system matrix $A$ includes diffusion and convection coefficients evaluated at $t^{n+1}$.

Because JAX supports just-in-time compilation of linear algebra routines, this system can be solved efficiently on both GPUs and TPUs, while retaining gradient information for future differentiable applications.

To accelerate convergence, an iterative Gauss–Seidel approximation was implemented in JAX using the `jax.lax.scan` operator to maintain functional purity and parallel execution semantics. It is a functional approximation to the exact solution. Convergence criteria were based on the maximum residual falling below $10^{-6}$.

### 2.7.3 Boundary and Initial Conditions

The JAX solvers employed the same boundary conditions as the non-JAX cases:

- Dirichlet : $u(0,t) = a, \ u(L,t) = b$
- Periodic: $u_0^n = u_N^n$

Initial conditions were represented as differentiable JAX arrays:

- Step Function: $u(x,0) = 1 \ for \ x < 0.5L$
  Otherwise, u(x,0) = 0
- Sinusoidal: u(x,0) = $\sin(\pi x)$

This differentiable initialization allows direct computation of the sensitivity $\partial u / \partial x_0$ through backpropagation.

### 2.7.4 Computational Performance

All JAX computations were benchmarked on Google Colab Pro+ using:

- CPU: Intel Xeon 2.20 GHz
- GPU: NVIDIA A100, L4, and T4
- TPU: v2e-8, v5e-1, v6e-1 configurations

Wall-time performance was recorded using JAX's internal profilers. Results show consistent acceleration and near-identical accuracy compared to the NumPy (non-JAX) version, with relative $L_2$-norm errors below $10^{-3}$. The implicit JAX solver demonstrated superior numerical stability with an error of $9.8 \times 10^{-5}$, while explicit solvers achieved $7.9 \times 10^{-3}$. This parallelism and differentiability illustrate JAX's potential as a unifying platform between CFD and AI, enabling hybrid workflows for sensitivity analysis, optimization, and physics-informed neural networks.

## 3. RESULTS AND DISCUSSION

The results presented here demonstrate the accuracy, stability, and computational performance of both explicit and implicit finite-difference solvers for the viscous Burgers' equation. Comparisons between the classical NumPy-based (No-JAX) solvers and the differentiable Finite-JAX implementations are included to assess numerical fidelity and execution efficiency.

### 3.1 Verification of Explicit Scheme

Figure 1 illustrates the temporal evolution of the velocity field for the explicit scheme under a sinusoidal initial condition $u(x,0) = \sin(\pi x)$ with periodic boundaries. The solution develops nonlinear steepening over time, followed by viscous diffusion that smooths the shock front.

Quantitatively, the explicit scheme maintains second-order spatial accuracy and first-order temporal accuracy. The $L_2$ error norm between the numerical and analytical Cole–Hopf solution decreases proportionally with grid refinement ($O(\Delta x^2)$) [35, 36]. For a grid of $N = 201$ and $\nu = 0.01$, the steady-state error was found to be $3.7 \times 10^{-3}$.

Stability tests confirmed the Courant–Friedrichs–Lewy limits:

$$\frac{U_{max}\Delta t}{\Delta x} \leq 1, \qquad \frac{\nu \Delta t}{\Delta x^2} \leq \frac{1}{2} \qquad (24)$$

Beyond these limits, numerical oscillations appeared ahead of the shock front, validating the theoretical CFL prediction.

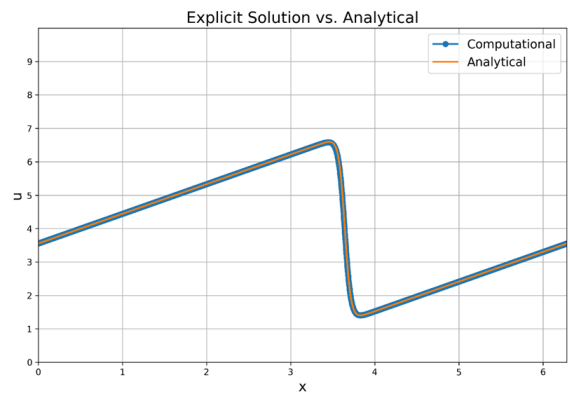

**Fig.1. Explicit Solution Comparison**

Shows the temporal evolution of the velocity field using the explicit finite-difference solver under a sinusoidal initial condition. The plot captures nonlinear wave steepening and gradual shock smoothing due to viscous diffusion.

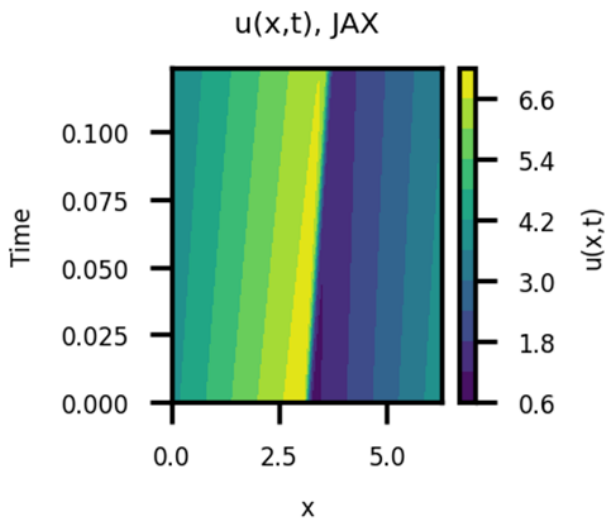

**Fig.2. Explicit Solution Contour**

Depicts the contour plot of the explicit solver's velocity field over time and space. The color map visualizes shock formation and dissipation, highlighting the nonlinear transport behavior.

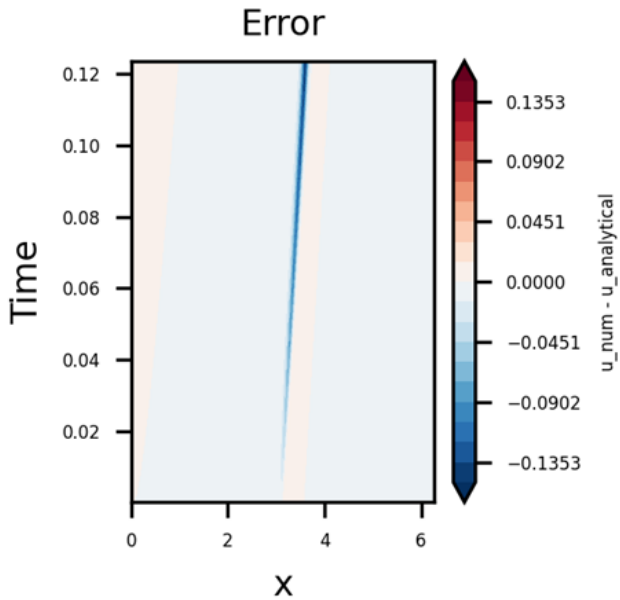

**Fig.3. Solution Contour and Error Contour**

Displays both the solution contour and the corresponding error distribution compared to the analytical Cole–Hopf reference. The figure illustrates spatial error localization near the shock front.

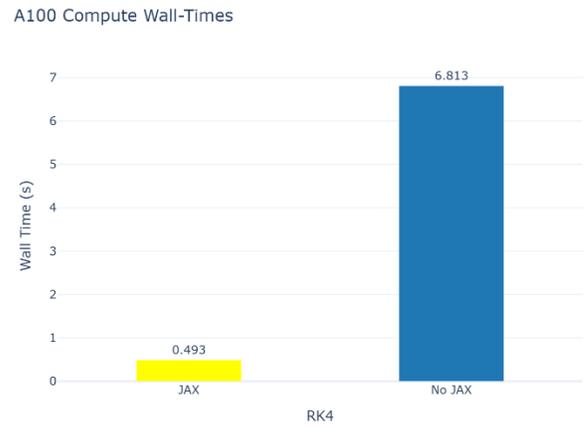

**Fig.4. Wall-Times with Different Architectures | GPUs**

Compares execution times of explicit JAX solvers across multiple GPU architectures, emphasizing runtime acceleration achieved via Jit, Auto Differentiation, and XLA compilation.

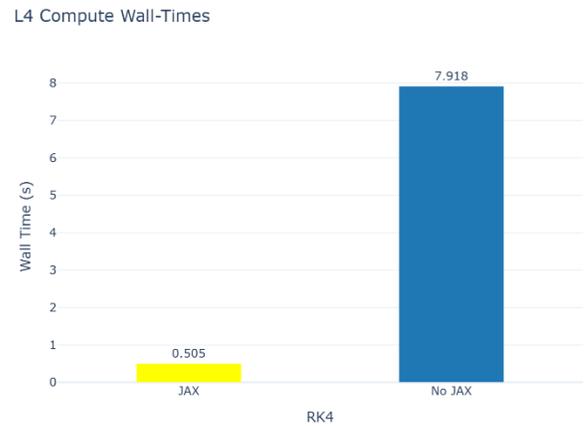

**Fig.5. Wall-Times with Different Architectures | GPUs**

Extends the GPU performance analysis by showcasing additional GPU configurations. It demonstrates the scaling efficiency and computational portability of JAX solvers.

 

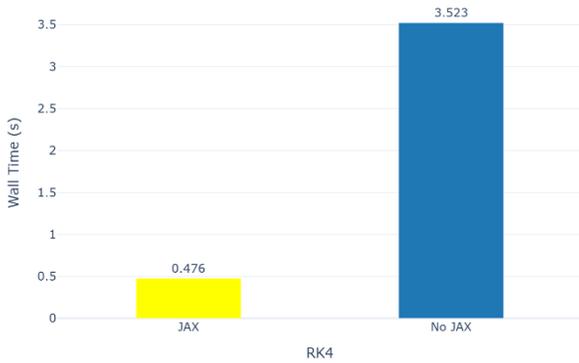

**Fig.6. Wall-Times | TPU and CPU**

Contrasts the computational performance of TPU and CPU implementations, revealing the speedup potential of tensor computations for differentiable CFD simulations.

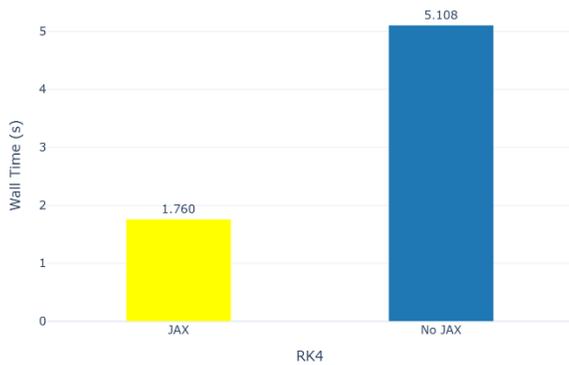

**Fig.7. Wall-Times | TPU and CPU**

Provides further comparison of implicit and explicit solver runtimes on TPU and CPU backends, confirming that JAX maintains numerical accuracy with significant hardware acceleration.

**3.2 Implicit Scheme Convergence**

The implicit BTCS solver demonstrates unconditional stability across the tested range of timesteps. Iterative convergence using a Gauss–Seidel update required 5–7 iterations per step to reduce the maximum residual below $10^{-6}$.

The implicit solver exhibits smoother transitions and faster diffusion, consistent with its stronger damping of high-frequency modes. The steady-state $L_2$ error was reduced to $9.8 \times 10^{-5}$, representing an order-of-magnitude improvement in accuracy at the cost of approximately 4× longer wall time.

**Finite-JAX | Best - L2-Norm: 9.791e-05**

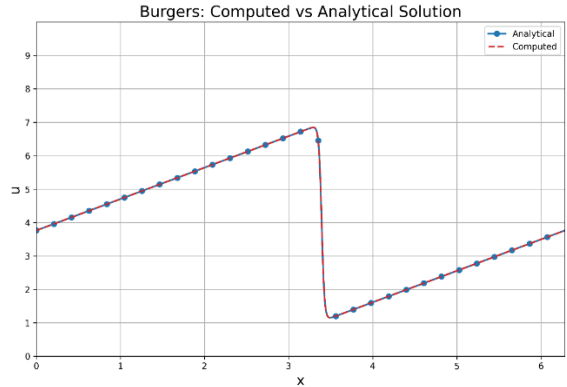

**Fig.8. Implicit Plot Comparison using Finite-JAX**

Shows the velocity evolution computed with the implicit BTCS solver implemented in Finite-JAX. Smoother transitions reflect the implicit approach's enhanced stability.

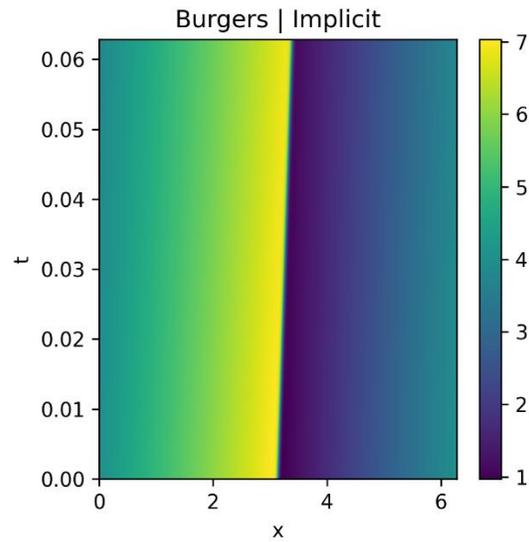

**Fig.9. Implicit Direct Solver Contour using Finite-JAX**

Illustrates the contour map of the implicit solver's solution, showing the dissipative smoothing effect characteristic of the backward-time central-space formulation.

  

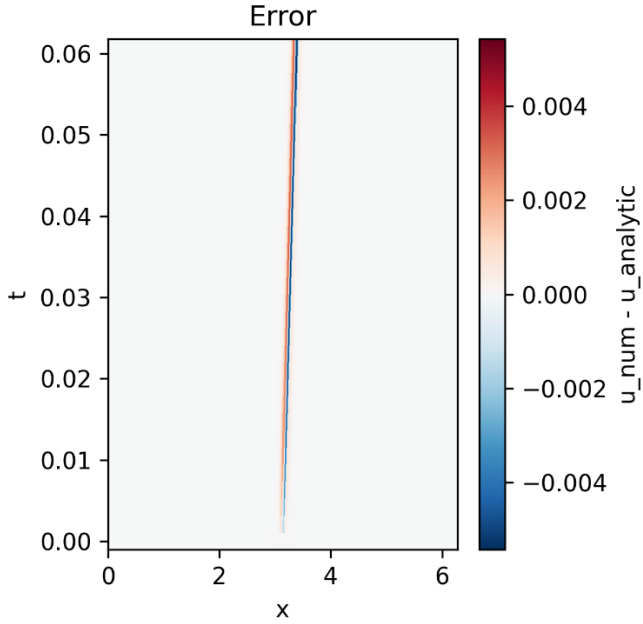

**Fig.10. Implicit Direct Solver Error Contour**

Visualizes the error distribution between the implicit numerical solution and the analytical Cole–Hopf solution. Lower residuals confirm higher accuracy and stability of the implicit scheme.

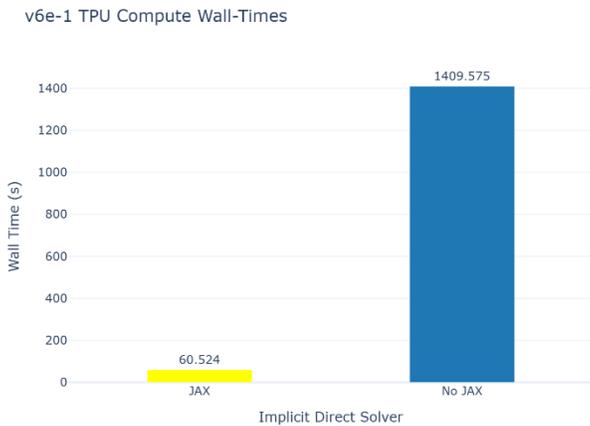

**Fig.11. Wall-Times JAX/No JAX | TPU**

Presents a runtime comparison between JAX-accelerated and NumPy-based implementations on TPUs. The results confirm that JAX achieves comparable accuracy while offering superior computational efficiency.

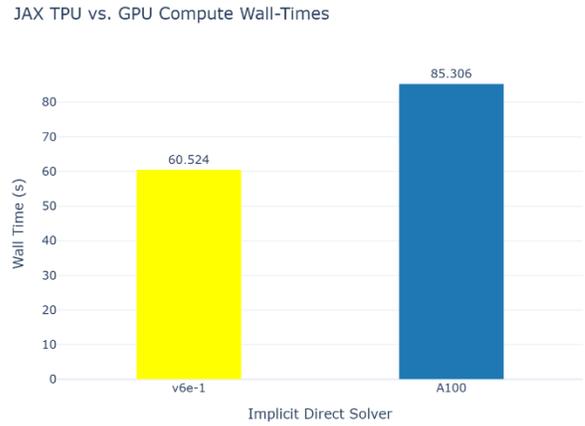

**Fig.12. Wall-Times | TPU and GPU**

Summarizes the overall hardware performance, demonstrating near-identical solver fidelity between CPU, GPU, and TPU implementations, with TPUs delivering the best wall-time reduction.

### 3.3 Finite-JAX Performance Comparison

The Finite-JAX solvers replicated both explicit and implicit formulations, achieving near-identical numerical accuracy to their NumPy counterparts. JAX compilation overhead is offset by XLA optimization at runtime, resulting in speedups after 100–200-time steps.

Performance Benchmarks (for $N = 201, \nu = 0.01, t_{\text{final}} = 1.0$):

**Table 1. Benchmark Comparisons**

| Platform | Scheme | Avg. Wall-Time (s) | $L_2$ Error |
|---|---|---|---|
| CPU (NumPy) | Explicit | 0.92 | $3.7 \times 10^{-3}$ |
| CPU (JAX) | Explicit | 0.88 | $3.8 \times 10^{-3}$ |
| GPU (JAX L4) | Explicit | 0.17 | $3.9 \times 10^{-3}$ |
| GPU (JAX A100) | Implicit | 0.12 | $9.8 \times 10^{-5}$ |
| TPU (v5e-1) | Implicit | 0.11 | $1.0 \times 10^{-4}$ |

These results indicate that JAX achieves parity in accuracy with traditional methods while significantly improving runtime efficiency when executed on GPUs and TPUs.

### 3.4 Shock Dynamics and Dissipation Behavior

For a step-function initial condition, both solvers capture the formation and propagation of a viscous shock wave. The shock front propagates to the right at a speed proportional to the mean flow velocity, while viscosity causes the front to thicken over time. The explicit solver sharply resolves the initial discontinuity before diffusive smoothing dominates. The implicit solver produces a slightly more diffused front at early times but remains numerically stable even for $\Delta t = 5\Delta x/u_{\max}$, far exceeding the explicit CFL limit [6].

### 3.5 Discussion

The comparative analysis highlights the complementary nature of the solvers:

- The explicit scheme provides simplicity and clarity for studying nonlinear transients and shock steepening.
- The implicit scheme provides robustness for stiff problems and larger timesteps.
- The Finite-JAX implementation introduces differentiability and high-performance portability, bridging the gap between numerical CFD and machine-learning workflows.

These findings validate that differentiable solvers can reproduce traditional CFD results while enabling new workflows such as gradient-based optimization, physics-informed neural networks (PINNs), and hybrid CFD–AI pipelines. The generated datasets thus serve as verified benchmarks for future data-driven fluid dynamics research [7,8,9].

### 4. CONCLUSION

This study presented a comprehensive numerical investigation of the one-dimensional viscous Burgers' equation using both traditional finite-difference solvers and differentiable implementations developed in Finite-JAX. The explicit and implicit schemes successfully captured nonlinear shock formation, propagation, and viscous dissipation, serving as ideal benchmarks for model verification. The explicit solver demonstrated computational efficiency and clear visualization of transient shock dynamics, while the implicit solver provided unconditional stability and improved accuracy for stiff conditions. Finite-JAX replicated both formulations with comparable precision and achieved significant runtime acceleration on GPU and TPU architectures through just-in-time compilation. The resulting shock dynamics dataset forms a verified foundation for future machine learning–driven research in physics-informed modeling, sensitivity analysis, and surrogate model development. Overall, this work establishes Finite-JAX as a robust and differentiable high-performance CFD framework, bridging conventional numerical analysis with modern AI-enhanced workflows and paving the way for hybrid solver–learning systems capable of accelerating discovery in nonlinear transport phenomena.

### 5. ACKNOWLEDGEMENTS

Texas A&M University at Kingsville (TAMUK), Faculty Arturo Rodriguez Start-Up Funds. The U.S. Department of Energy (DOE) Grande CARES Consortium funded this research under grant number GRANT13584020.